\documentclass[conference]{IEEEtran}
\IEEEoverridecommandlockouts
\usepackage{multirow}
\usepackage{cite}
\usepackage{amsmath,amssymb,amsfonts}
\usepackage{algorithmic}
\usepackage{graphicx}
\usepackage{textcomp}
\usepackage{xcolor}
\usepackage{comment}
\usepackage{diagbox} 
\newcommand{\fref}[1]{Fig.\ {\ref{#1}}}
\newcommand{\tref}[1]{TABLE\ {\ref{#1}}}
\def\BibTeX{{\rm B\kern-.05em{\sc i\kern-.025em b}\kern-.08em
    T\kern-.1667em\lower.7ex\hbox{E}\kern-.125emX}}
\begin{document}

\title{\huge Passive Reactance Compensation for Shape-Reconfigurable Wireless Power Transfer Surfaces}

\author{\IEEEauthorblockN{Riku Kobayashi, Yoshihiro Kawahara, and Takuya Sasatani}
\IEEEauthorblockA{Graduate School of Engineering, The University of Tokyo, Bunkyo-ku, Japan \\
\{riku, kawahara, sasatani\}@akg.t.u-tokyo.ac.jp}}

\maketitle

\newcommand\blfootnote[1]{
    \begingroup
    \renewcommand\thefootnote{}\footnote{#1}
    \addtocounter{footnote}{-1}
    \endgroup
}
\blfootnote{© 2025 IEEE. Personal use of this material is permitted. Permission from IEEE must be obtained for all other uses, in any current or future media, including reprinting/republishing this material for advertising or promotional purposes, creating new collective works, for resale or redistribution to servers or lists, or reuse of any copyrighted component of this work in other works.

This is the author-submitted version of the manuscript accepted for future publication at the 2025 IEEE Wireless Power Technology Conference and Expo (WPTCE). The final published version may differ from this manuscript.}

\begin{abstract}

The powering range of wireless power transfer (WPT) systems is typically confined to areas close to the transmitter. Shape-reconfigurable two-dimensional (2-D) relay resonator arrays have been developed to extend this range, offering greater deployment flexibility. However, these arrays encounter challenges due to cross-coupling among adjacent resonators, which detune system impedance and create power dead zones. This issue often necessitates active components such as receiver position tracking, increasing system overhead. This study introduces a passive reactance compensation mechanism that counteracts detuning effects, enabling the simultaneous activation of all resonators at a fixed operating frequency, regardless of the array's shape, thus providing a consistent charging area. The key innovation involves mechanically appending reactance elements to neutralize detuning caused by inductive coupling, facilitating hassle-free resonator reconfiguration without requiring prior knowledge. Our experiments demonstrate the elimination of dead zones with multiple configurations, boosting the minimum power transfer efficiency from 3.0\% to 56.8\%.

\end{abstract}


\section{Introduction}


Wireless power transfer (WPT) using magnetic resonant coupling efficiently powers electronic devices even when there is misalignment between the transmitter and receiver, benefiting applications such as mobile devices~\cite{mobile_devices} and wireless sensor systems in free-moving animal experiments~\cite{optgenetics}. However, effective power transfer is typically restricted to the area near the transmitter. To overcome these spatial limitations and accommodate diverse charging needs, researchers have developed two-dimensional (2-D) transmitter (TX) resonator arrays~\cite{SRCA}. Recent advancements have led to shape-reconfigurable 2-D resonator arrays, which distribute energy from the TX resonator via relay resonators, offering greater deployment flexibility in various environments and configurations~\cite{alvus}.

\begin{figure}[t]
\centerline{\includegraphics[width=1.0\linewidth]{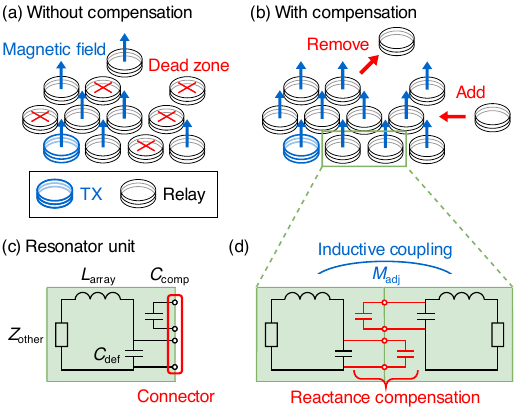}}
\caption{Concept and principle of the proposed compensation mechanism: (a) Simply arranging $LC$ resonators in a 2-D array leads to dead zones due to inductive coupling between resonators. (b) The proposed approach compensates for detuning caused by coupling via mechanical connections, enabling a consistent charging range across the array. (c) Basic schematic of the proposed passive reactance compensation mechanism, where $L_{\rm array}$ represents the inductance, $C_{\rm def}$ is the series capacitance, $C_{\rm comp}$ is the compensation capacitance, and $Z_{\rm other}$ denotes the remaining circuit impedance. (d) Inductive coupling between adjacent resonators, characterized by mutual inductance $M_{\rm adj}$, is offset by a capacitor attached by the neighboring unit.}
\label{fig:1}
\end{figure}

\begin{figure*}[t]
\centerline{\includegraphics[width=1.0\linewidth]{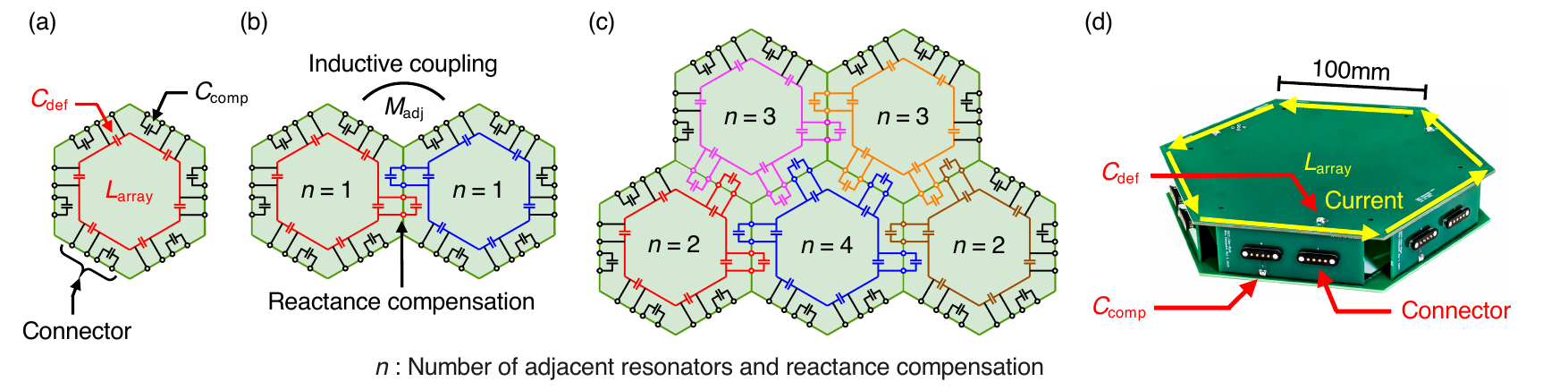}}
\caption{Overview of the proposed reactance compensation method. (a-c) Operating principles of the deformable resonator coil array are illustrated. The colored traces indicate the current loop forming the resonator, while $n$ denotes the number of adjacent resonators for each unit. Note that inductance symbols are omitted. (a) Schematic of a single resonator array unit, displaying each edge with a default capacitor $C_{\rm def}$ and a compensation capacitor $C_{\rm comp}$, which offsets the reactance of adjacent resonators. (b) When two resonator units are connected, impedance detuning from inductive coupling with mutual inductance $M_{\rm adj}$ is compensated via the attached compensation capacitance. (c) This principle extends to configurations with more resonator units, as the number of inductive couplings and connected compensation capacitors naturally align. (d) Implemented prototype of the resonator unit employing the proposed reactance compensation method.}
\label{fig:2}
\end{figure*}

A major challenge with these 2-D resonator arrays is the prevention of simultaneous activation of multiple adjacent resonators. Such activations can induce cross-coupling due to inductive interactions between resonators, which detune the system's impedance and resonance, creating power dead zones even when the resonators are active, as shown in \fref{fig:1}(a). Typical solutions involve continuously tracking the receiver's position to selectively activate the nearest transmitter\cite{SRCA} or activating a specific set of relay resonators to minimize interference\cite{alvus}. However, these methods increase overhead and complexity for state detection and control, particularly when charging multiple or dynamically moving targets.

This study presents a passive reactance compensation method that neutralizes the detuning effects caused by adjacent resonators. It enables the simultaneous activation of all array resonators--including a TX resonator and multiple relay resonators--at a fixed frequency, regardless of their shape, as shown in \fref{fig:1}(b). This advancement facilitates a shape-reconfigurable wireless power transfer surface capable of charging multiple dynamic devices within its range without active detection or control. A key advantage of this approach is that it requires no prior knowledge or adjustments when the array is deformed, ensuring ease of deployment and use.

For homogeneous resonator arrays, where all units are identical in shape (\textit{e.g.}, hexagonal or square), detuning from cross-coupling can be mitigated by integrating a reactance element into each resonator, proportional to the number of its adjacent resonators~\cite{hercules}. Our approach utilizes this concept, demonstrating that compensation can occur by mechanically and mutually attaching capacitive elements to neighboring units to offset the inductive coupling, as shown in \fref{fig:1}(c,d). Consequently, each instance of detuning can be naturally compensated as it arises, allowing for consistent magnetic field coverage regardless of the array's overall shape. We demonstrate this approach and confirm that it effectively creates shape-reconfigurable wireless power transfer surfaces free of dead zones, thus eliminating the need for receiver detection or transmitter control.

\section{Principle of Detuning and Compensation}\label{sec:principle}
Arranging relay ($LC$) resonators in a 2-D resonator array does not inherently produce a consistent magnetic field across the array surface. This arrangement often results in dead zones with weak magnetic fields and low power transfer efficiency, as illustrated in \fref{fig:1}(a). The issue arises from the coupling between adjacent resonators, which introduces a reactance that detunes the impedance of each resonator unit. As the number of resonators increases, this detuning effect escalates in complexity and severity, making it exceedingly difficult to achieve a uniform magnetic field across the entire resonator array surface.

This study assumes a homogeneous resonator array configuration, where all units have identical shapes (\textit{e.g.}, hexagonal or square). Previous research has shown that by adjusting the capacitive impedance
$Z_{\rm C}$ of each unit according to the equation
\begin{equation}
Z_{\rm C}=-j\omega_0(L_{\rm array}+nM_{\rm adj}),
\label{eq:1}
\end{equation}
the periodicity of the array can be leveraged to compensate for the detuning factor caused by adjacent inductive coupling~\cite{hercules}. In this equation, $n$ is the number of adjacent resonators, $L_{\rm array}$ represents the self-inductance of each coil, $M_{\rm adj}$ denotes the mutual coupling between adjacent resonators, and $\omega_0$ is the operating angular frequency. This adjustment theoretically ensures consistent current flow across all resonator units, enabling a uniform magnetic field and efficient power transfer throughout the array surface.

The challenge lies in seamlessly achieving the impedance condition formulated in \eqref{eq:1}. The required impedance varies with the number of adjacent resonators $n$, which naturally changes with array deformation. Consequently, a straightforward selection of passive components cannot dynamically fulfill this condition. Our approach cleverly utilizes the connections with adjacent resonators—the main source of detuning—to mechanically append reactance components and achieve the required impedance condition in synchronization with changes in $n$. The following section discusses the analysis and mechanism supporting this approach.

\section{Passive Reactance Compensation Method}
In this section, we introduce a passive reactance compensation method designed to maintain the impedance condition formulated in \eqref{eq:1}. For the sake of simplification, we assume hexagonal resonators, as utilized in prior studies~\cite{alvus,hercules}, although the analysis can potentially be extended to other geometries, such as square coils.

The core concept involves creating an interface at each edge of the array resonators that can externally adjust the resonator unit impedance by $-j\omega_0 M_{\rm adj}$. This adjustment cancels out the detuning reactance introduced by a single adjacent resonator, as specified in \eqref{eq:1}. Implementation requires two types of capacitors: (i) default capacitors $C_{\rm def}$, which are assigned to each edge of the resonator to define its default resonant frequency, and (ii) compensation capacitors $C_{\rm comp}$, which counteract detuning effects due to resonator adjacency. \fref{fig:2}(a-c) illustrates the placement and integration of these capacitors within the resonator arrangement.

The value of the default capacitors $C_{\rm def}$ can be derived from the series resonance condition $\omega = 1/\sqrt{LC}$, with a modification to account for distribution on six edges across the coil path:
\begin{equation}
C_{\rm def}=\frac{6}{\omega_0^2 L_{\rm array}},
\label{eq:2}
\end{equation}
where $L_{\rm array}$ is the inductance of the resonator coil. When there are $n$ adjacent resonators and $n$ compensation capacitors $C_{\rm comp}$ are attached, the condition in \eqref{eq:1} can be rewritten as:
\begin{equation}
\frac{6-n}{C_{\rm def}}+\frac{n}{C_{\rm def}+C_{\rm comp}}=\omega_0^2L_{\rm array}(1+nk_{\rm adj}),
\label{eq:3}
\end{equation}
where $k_{\rm adj}$ is the coupling coefficient between array resonators ($M_{\rm adj} = k_{\rm adj}L_{\rm array}$). The necessary value for $C_{\rm comp}$ can be calculated using \eqref{eq:2} and \eqref{eq:3}:
\begin{equation}
C_{\rm comp}=\frac{-36k_{\rm adj}}{\omega_0^2L_{\rm array}(1+6k_{\rm adj})} 
\label{eq:4}
\end{equation}
Notably, both $C_{\rm def}$ and $C_{\rm comp}$ are independent of $n$, allowing the impedance condition specified in \eqref{eq:1} to be maintained using only static passive capacitor components.

\fref{fig:2}(d) shows the prototype implementation of the resonator unit. Coil traces are located along the edges of the hexagonal printed circuit board~(PCB) with a side length of $l_{\rm array}=100~\mathrm{mm}$, forming a two-turn inductor $L_{\rm array}$. The default capacitors $C_{\rm def}$ are integrated along the coil trace, while the compensation capacitors $C_{\rm comp}$ are positioned at the edges of the resonator module and attached via magnetic connectors when adjacent resonators are present.

\section{Simulations and Experiments}
We evaluated our approach through two-port measurements using a vector network analyzer (VNA) and SPICE simulations. Initially, we measured the frequency response of the resonator array in various deformed configurations to verify that the reactance compensation method effectively mitigates the detuning caused by coupling between resonators. Subsequently, we measured and simulated the power transfer efficiency over each resonator coil to confirm that our approach successfully eliminates dead zones.

These evaluations were conducted using four resonator units, as shown in \fref{fig:2}(d). Measured and defined parameters used in the evaluation, including the operating frequency $f_0$ (where the corresponding angular frequency is given by $\omega_0=2\pi f_0$) and circuit specifications, are listed in Table~\ref{tab:1}. Note that minimal manual adjustments were made when defining capacitor values to account for non-ideal parameters, such as parasitic capacitance.

\begin{table}[t]
\caption{Resonator array parameters}
\begin{center}
\begin{tabular}{l|l|l}
\textbf{Description}&\textbf{Symbol}&\textbf{Value} \\
\hline
Operating frequency & $f_0$ & $13.56\,\mathrm{MHz}$ \\
\hline
Array resonator inductance & $L_{\rm array}$ & $2.09\,\mathrm{uH}$ \\
\hline
Coupling coefficient between
& \multirow{2}{*}{$k_{\rm adj}$} & \multirow{2}{*}{$-0.0960$} \\
adjacent array resonators & & \\
\hline
Default capacitance & $C_{\rm def}$ & $396\,\mathrm{pF}$ \\
\hline
Compensation capacitance & $C_{\rm comp}$ & $596\,\mathrm{pF}$ \\
\hline
Array coil Q-factor& $Q_{\rm array}$ & $220$  \\
\hline
Side length of resonators & $l_{\rm array}$ & $100\,\mathrm{mm}$ \\
\end{tabular}
\vspace{5mm}
\label{tab:1}
\end{center}

\caption{Receiver coil parameters}
\begin{center}
\begin{tabular}{l|l|l}
\textbf{Description}&\textbf{Symbol}&\textbf{Value} \\
\hline
Receiver inductance& $L_{\rm RX}$ & $3.11\,\mathrm{uH}$ \\
\hline
Coupling coefficient between
& \multirow{2}{*}{$k_{\rm RX}$} & \multirow{2}{*}{$0.0684$} \\
the receiver and the nearest array resonator & & \\
\hline
Receiver capacitance& $C_{\rm RX}$ & $44\,\mathrm{pF}$ \\
\hline
Receiver Q-factor& $Q_{\rm RX}$ & $275$ \\
\hline
Receiver diameter & $d_{\rm RX}$ & $50\,\mathrm{mm}$ \\
\end{tabular}
\label{tab:2}
\end{center}
\end{table}

\begin{figure}[t]
\centerline{\includegraphics[width=1.0\linewidth]{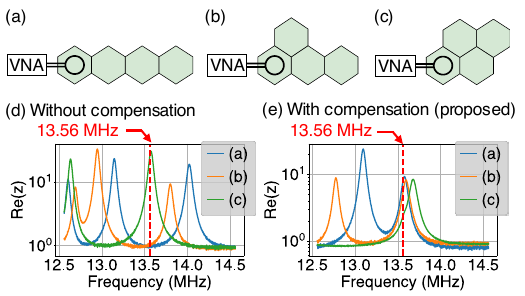}}
\caption{Frequency response measurements of the proposed deformable resonator array. (a-c) resonator array shapes used for measuring frequency response, with the black circle indicating the external measurement coil. The measured frequency response is shown (d) without compensation and (e) with the proposed compensation method. The resonant frequency of a single transmitter resonator is 13.56~MHz.}
\label{fig:3}
\end{figure}

\begin{figure}[t]
\centerline{\includegraphics[width=1.0\linewidth]{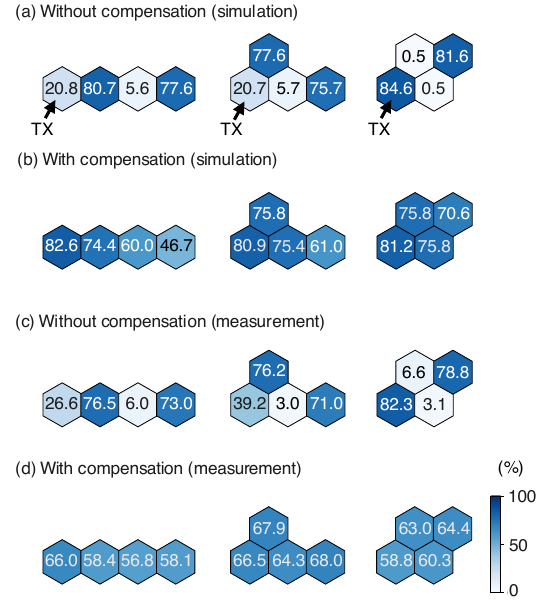}}
\caption{
Simulated and measured power transfer efficiency using the proposed shape-reconfigurable wireless power transfer surface. Results are presented with and without the proposed compensation method to illustrate its impact. ``TX'' indicates the resonator connected to the power source. (a) Simulation results without compensation. (b) Simulation results with the proposed compensation. (c) Measurement results without compensation. (d) Measurement results with the proposed compensation method.
}
\label{fig:4}
\end{figure}

\subsection{Frequency Response of the deformed array}\label{A}
We measured the frequency response of the array while deforming it into the three shapes shown in \fref{fig:3}(a-c). These measurements were conducted using an external coil connected to a VNA to capture the real part of the impedance. For comparison, measurements were performed both with and without implementing our passive reactance compensation method. It is expected that the resonant frequency varies less when using the compensation.

\fref{fig:3}(d) presents the results obtained without compensation, while \fref{fig:3}(e) displays the results with the proposed compensation method. These findings indicate that the reactance compensation mechanism maintains a constant resonant frequency, regardless of the resonators' arrangement. This demonstrates that the inherent detuning observed in \fref{fig:3}(d) is effectively compensated through our approach.

\subsection{Power Transfer Efficiency of the deformed array}\label{B}

Next, we measured the power transfer efficiency of the deformed resonator arrays configured as shown in \fref{fig:3}(a-c). One of the resonators in each resonator array configuration was connected to the power source, and a circular receiver coil was positioned at the center of each resonator unit at a distance of 1 cm. Parameters for the receiver, including the inductance $L_{\rm RX}$, coupling coefficient to the nearest resonator $k_{\rm RX}$, capacitance $C_{\rm RX}$, Q-factor $Q_{\rm RX}$, and diameter $d_{\rm RX}$, are listed in Table~\ref{tab:2}. For measurements, we calculated the maximum power transfer efficiency at the resonant frequency $f_0$ using data from measured two-port networks~\cite{zargham2012maximum}. Efficiency values for simulations were derived from circuit simulations integrating measured circuit parameters.

\fref{fig:4}(a,b) illustrates the simulated efficiency without and with the proposed reactance compensation approach. \fref{fig:4}(c,d) depicts the measured efficiency without and with the proposed compensation method. Each efficiency value in \fref{fig:4}(a-d) represents the power transfer efficiency when the receiver is placed on the resonator at the corresponding position. These figures demonstrate that dead zones, particularly those with efficiency below 10\%, are eliminated using our compensation approach. Notably, the minimum efficiency improves in all configurations. The minimum efficiency across all configurations increased from 0.5\% to 46.7\% in simulations and from 3.0\% to 56.8\% in measurements, confirming the effectiveness of our approach.

\section{Conclusion}
This study introduced a passive reactance compensation mechanism that enhances shape-reconfigurable wireless power transfer surfaces by addressing detuning issues in 2-D resonator arrays. Our mechanism maintains stable impedance conditions through cleverly integrated capacitors, ensuring robust power transfer across various shapes and configurations without active control. Our results demonstrate significant improvements in power transfer efficiency, with minimum efficiencies increasing from 0.5\% to 46.7\% in simulations and from 3.0\% to 56.8\% in measurements. Eliminating dead zones, especially those with efficiencies below 10\%, confirms the effectiveness of our approach. This advancement allows for quickly deploying WPT surfaces that support dynamic device positioning, paving the way for more adaptable and efficient wireless charging solutions. Potential future work includes large-scale implementations and applying this approach to receivers for deformable devices~\cite{kadomoto2021toward}.

\section*{Acknowledgment}
This work was supported by JSPS KAKENHI Grant Number 23K28068, 23H03378, JST ACT-X Grant Number JPMJAX190F, and Hirose Foundation.

\bibliographystyle{IEEEtran}
\bibliography{refer} 

\end{document}